\documentclass[11pt]{article}

\usepackage[margin=1in]{geometry}
\usepackage{amsmath}
\usepackage{amssymb}
\usepackage{caption}
\usepackage{subcaption}
\usepackage{graphicx}
\usepackage{tabularx}
\usepackage{booktabs}
\usepackage{hyperref}
\usepackage{natbib}
\usepackage{xcolor}
\usepackage{wrapfig}
\usepackage{placeins}
\usepackage{algorithm}
\usepackage{algorithmic}
\newcommand{\mathbbm}[1]{\mathbf{#1}}
\usepackage{enumitem}
\usepackage{tcolorbox}

\title{Grounding Social Perception in Intuitive Physics}

\author{%
Lance Ying\textsuperscript{1,2,*},
Aydan Y. Huang\textsuperscript{3,*},
Aviv Netanyahu\textsuperscript{1,*},
Andrei Barbu\textsuperscript{4},
Boris Katz\textsuperscript{1}\\[0.3em]
Joshua B. Tenenbaum\textsuperscript{1},
Tianmin Shu\textsuperscript{3,\dag}%
\\[1.5em]
{\small\itshape \textsuperscript{1}Massachusetts Institute of Technology \quad
\textsuperscript{2}Harvard University}\\
{\small\itshape \textsuperscript{3}Johns Hopkins University \quad
\textsuperscript{4}Amazon}%
}

\date{}

\begin{document}

\maketitle

\renewcommand{\thefootnote}{\fnsymbol{footnote}}
\footnotetext[1]{These authors contributed equally to this work.}
\footnotetext[2]{Corresponding author. E-mail: tianmin.shu@jhu.edu}
\renewcommand{\thefootnote}{\arabic{footnote}}
\setcounter{footnote}{0}

\begin{abstract}
People infer rich social information from others' actions. These inferences are often constrained by the physical world: what agents can do, what obstacles permit, and how the physical actions of agents causally change an environment and other agents' mental states and behavior. We propose that such rich social perception is more than visual pattern matching, but rather a reasoning process grounded in an integration of intuitive psychology with intuitive physics. To test this hypothesis, we introduced PHASE (PHysically grounded Abstract Social Events), a large dataset of procedurally generated animations, depicting physically simulated two-agent interactions on a 2D surface. Each animation follows the style of the Heider and Simmel movie, with systematic variation in environment geometry, object dynamics, agent capacities, goals, and relationships (friendly/adversarial/neutral). We then present a computational model, SIMPLE, a physics-grounded Bayesian inverse planning model that integrates planning, probabilistic planning, and physics simulation to infer agents' goals and relations from their trajectories. Our experimental results showed that SIMPLE achieved high accuracy and agreement with human judgments across diverse scenarios, while feedforward baseline models---including strong vision-language models---and physics-agnostic inverse planning failed to achieve human-level performance and did not align with human judgments. These results suggest that our model provides a computational account for how people understand physically grounded social scenes by inverting a generative model of physics and agents.
\end{abstract}

\vspace{0.5em}

\section{Introduction}

\begin{figure*}[t!]
    \centering
    \includegraphics[width=0.95\linewidth]{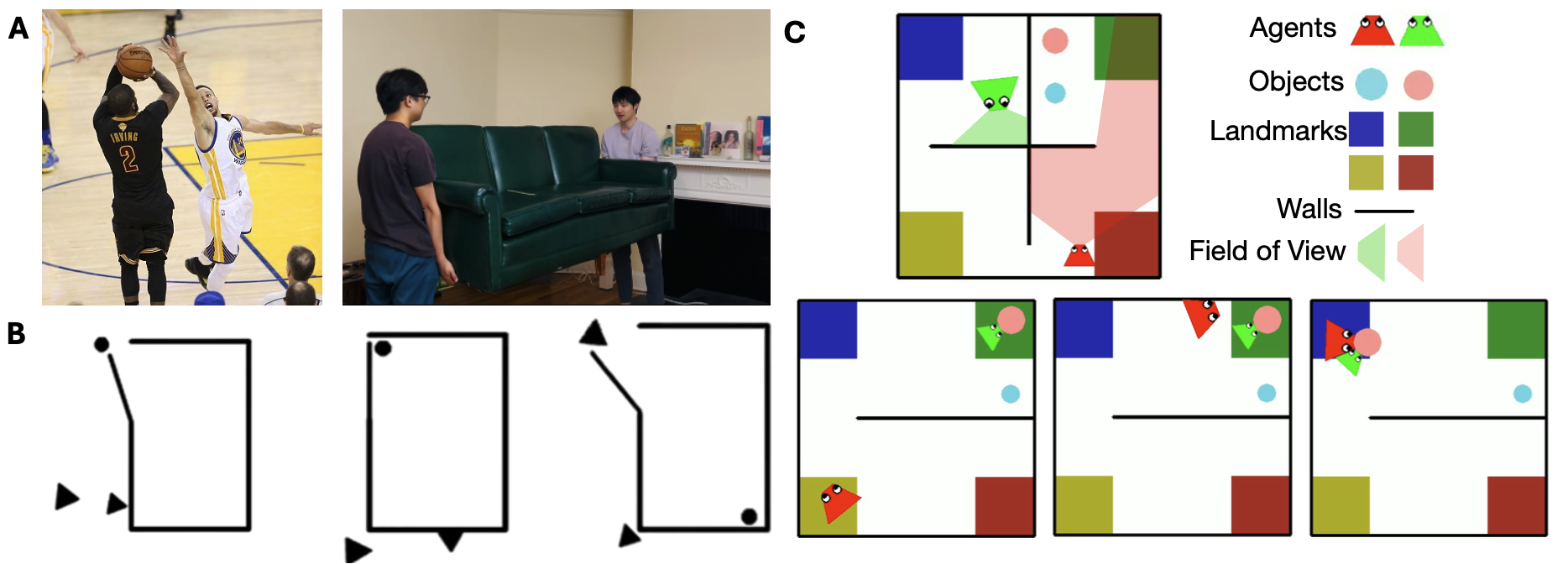}
    \caption{(\textbf{A}) Examples of real-life social interactions in physical environments: a basketball player trying to block an opponent from shooting a ball; two people carrying a couch together.
    (\textbf{B}) The classic Heider-Simmel animation abstracts such real-life interactions in animated displays of simple geometric shapes.
    (\textbf{C}) \textit{PHASE} models social interactions as physically grounded abstract events, where animated agents with limited fields of view move in a physics-based environment with objects, landmarks, and obstacles, enabling behaviors such as helping, hindering, or collaborating toward goals.}
    \label{fig:intro}
\end{figure*}

Human social perception is richly grounded in our understanding of physics. As illustrated in Figure~\ref{fig:intro}\textbf{A}, when we infer that one agent is chasing, blocking, or helping another, we rely not only on intuitive psychology (e.g., agents act purposefully) but also on intuitive physics (e.g., bodies move continuously, forces have effects, and obstacles constrain what is possible). Classic demonstrations such as the Heider and Simmel movie \citep{heider1944experimental} show that even movements of simple geometric shapes can elicit rich social narratives (as depicted in Figure~\ref{fig:intro}\textbf{B}). Critically, those anthropomorphic narratives of social behavior rely on viewers' interpretation of the physical interactions (e.g., contact, collision, and impeded motion). This raises a core cognitive science question: how do people infer rich social information from physically grounded observations?

There have been many computational accounts of social perception that capture aspects of rational action understanding. One prominent type of computational approaches is inverse planning \citep{baker2009action,baker2017rational, jara2020naive, ying2025language}, which posits that people assume others are rational planners that select the most optimal action to achieve their goal, minimizing costs and maximizing utilities. The observer can then invert this generative model to compute the most likely mental states that can best explain the observed action. However, existing models of inverse planning \citep{baker2009action,baker2017rational,jara2016naive,ullman2009help,zhi2020online} often abstract away physical dynamics, treating actions as symbolic state transitions. Conversely, recent end-to-end neural models have emerged as powerful approaches to model social perception by training on large corpora of diverse social scenes \citep{rabinowitz2018machine, hovaidi2018neural, malik2023relational}. Although these models can learn predictive representations from large corpora, they typically rely on visual features and also do not explicitly show sophisticated physical reasoning processes.

In this paper, we study how people flexibly interpret social scenes in physically grounded environments. We hypothesize that people can utilize an analysis-by-synthesis approach \citep{yuille2006vision} on these tasks: we possess a joint physical and social generative model of agents planning in our minds and invert this generative process to infer the most likely goal, relations, and physical parameters that best explain the observation. This view builds on prior work on infants' judgment of agents' goal-directed actions in physical environments \citep{liu2017ten,liu2022dangerous, saxe2005secret}. As prior work has demonstrated, people, starting from as early as 10 months old, can estimate the costs of actions for pursuing goals under various physical constraints, and reason about cost-utility trade-offs to interpret their behavior. Recent studies in neuroscience and computational cognitive science have also provided evidence that people's intuitive theory of psychology builds on their intuitive theory of physics \citep{liu2025physical}. In this work, we extend the prior work and operationalize this view with a dataset and a computational model.

Our main contributions are twofold. First, we introduce PHASE (PHysically grounded Abstract Social Events), a dataset of physically simulated two-agent interactions with controlled variation in scene geometry, physical dynamics, agent capacities, and social structure (goals and relationships), as illustrated Figure~\ref{fig:intro}\textbf{C}. Second, we propose SIMPLE (SIMulation, Planning, and Local Estimation), a physics-grounded Bayesian inverse planning model that combines a forward physics engine with probabilistic planning and inference over goals and relations. We show that incorporating physical dynamics improves goal and relationship inference relative to physics-agnostic inverse planning and end-to-end video baselines, and that the resulting inferences better align with human judgments.

\section{A Paradigm for Physically Grounded Social Perception}

\begin{figure*}[t!]
\centering
    \includegraphics[width=0.85\linewidth]{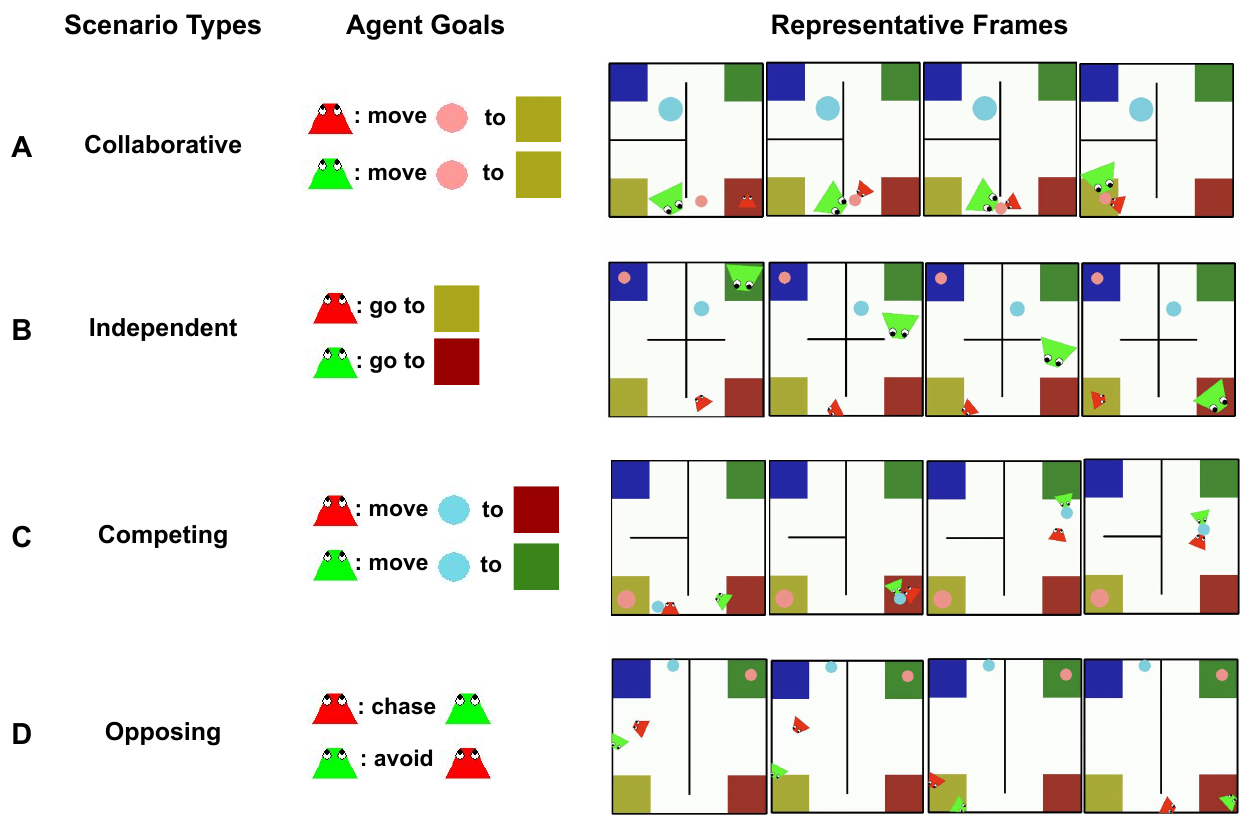}
    \caption{Example PHASE animations for different scenario types. A). In this collaborative scenario, agents collaborate to move the pink circle to the yellow landmark. The green agent is too large to move past the barrier. Therefore, the red agent first moves the circle to the left of the barrier and then carries it together with the green agent to the yellow landmark. B). Two agents move towards different landmarks with no interaction. C). In this competing scenario, the agents want to move the blue circle to different landmarks, but there is only one blue circle. They therefore pull the circle in different directions. D). In this opposing scenario, the red agent tries to chase the green agent, while the green agent tries to get away from the red agent. The videos for these examples can be found at \href{https://osf.io/fkp5m/}{https://osf.io/fkp5m/}. }
    \label{fig:examples}
\end{figure*}

Inspired by the seminal work by Heider and Simmel \citep{heider1944experimental}, we curated the PHASE dataset to study human social perception in abstract physically grounded scenes. The videos of the PHASE dataset feature two agents in the shape of trapezoids. The agents can vary in their sizes and strengths. They have limited fields of view. There are four landmarks in the scene, each of which is a square at one of the four corners. There are objects in the scene, represented as circles, which agents may or may not be able to move depending on their strengths.

\begin{table}[t!]
    \centering
    \small
    \begin{tabularx}{\columnwidth}{@{} l >{\raggedright\arraybackslash}X l @{}}
        \toprule
        \textbf{Scenario Type} & \textbf{Description} & \textbf{Relation} \\ \midrule
        Collaborative        & Two agents collaborate on a shared goal. & Friendly \\ \addlinespace
        Independent & The two agents have different, non-conflicting goals. & Neutral \\ \addlinespace
        Competing   & The two agents have conflicting goals about the same object. & Adversarial \\ \addlinespace
        Opposing    & The agents have opposite goals, where one agent's goal is to intentionally prevent the other agent from reaching its goal. & Adversarial \\ \bottomrule
    \end{tabularx}
    \caption{Types of agent goal scenarios, their descriptions, and the corresponding relations.}
    \label{tab:goal_scenarios}
\end{table}

The agent's objective is to move to a landmark, move an object to a landmark, move close to the other agent, or hinder a particular goal (e.g., keep an agent or an object away from a landmark). By permuting the objects and landmarks, we generate 36 possible goals in total for the goal space.

As each video contains two agents who may have the same or different goals, we classify the pair of agent goals in each video into 4 scenario goal types: collaborative, independent, competing, and opposing as defined in Table \ref{tab:goal_scenarios}. Figure~\ref{fig:examples} shows an example for each type. These scenario goal types map onto three relation labels: friendly, neutral, and adversarial.

\begin{figure}[t!]
    \centering
    \includegraphics[width=0.6\textwidth]{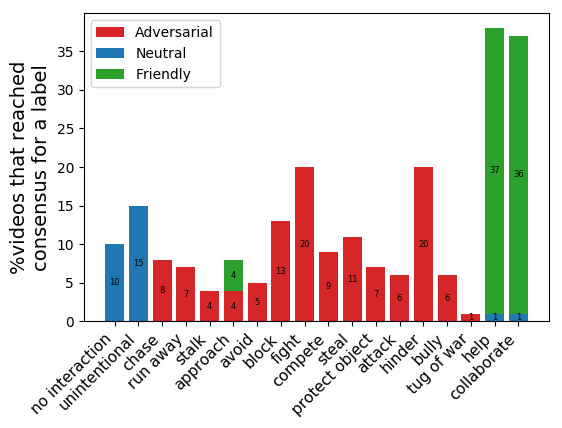}
    \caption{Consistent human responses showing how many videos (percentages) were assigned with an interaction category by at least 50\% of the participants who have watched the videos.}
    \label{fig:video_annotation}
\end{figure}

PHASE contains 500 videos of abstract social events. Each goal category has 47--129 examples. Relationship labels are distributed as 200 friendly, 192 adversarial, and 108 neutral videos. We split the dataset into 320 training, 80 validation, and 100 test videos. Our human experiment validated that people would consistently label the 100 test videos with 23 different types of social interactions, as shown in Figure~\ref{fig:video_annotation}.

The PHASE dataset covers diverse social interaction scenarios. We compiled a list of 23 interaction labels from prior work \citep{gao2011chasing,gordon2014authoring} and from free-form descriptions collected in a pilot study (see Appendix 1A). We then recruited 130 participants to annotate 100 PHASE videos. Our analysis shows that the 23 labels were used to describe at least one video; moreover, under a majority-vote criterion, PHASE videos covered 18 distinct interaction categories (Figure~\ref{fig:video_annotation}). In another human experiment, we found that the test videos in PHASE were indistinguishable from human-generated animations in the same environment through a controller (see Appendix 1B).

\begin{figure*}[t!]
    \centering

    \begin{subfigure}[t]{0.50\textwidth}
        \centering
        \setlength{\unitlength}{1pt}
        \begin{picture}(0,0)
            \put(-140,-18){\large\textbf{A}}
        \end{picture}
        \includegraphics[width=\linewidth]{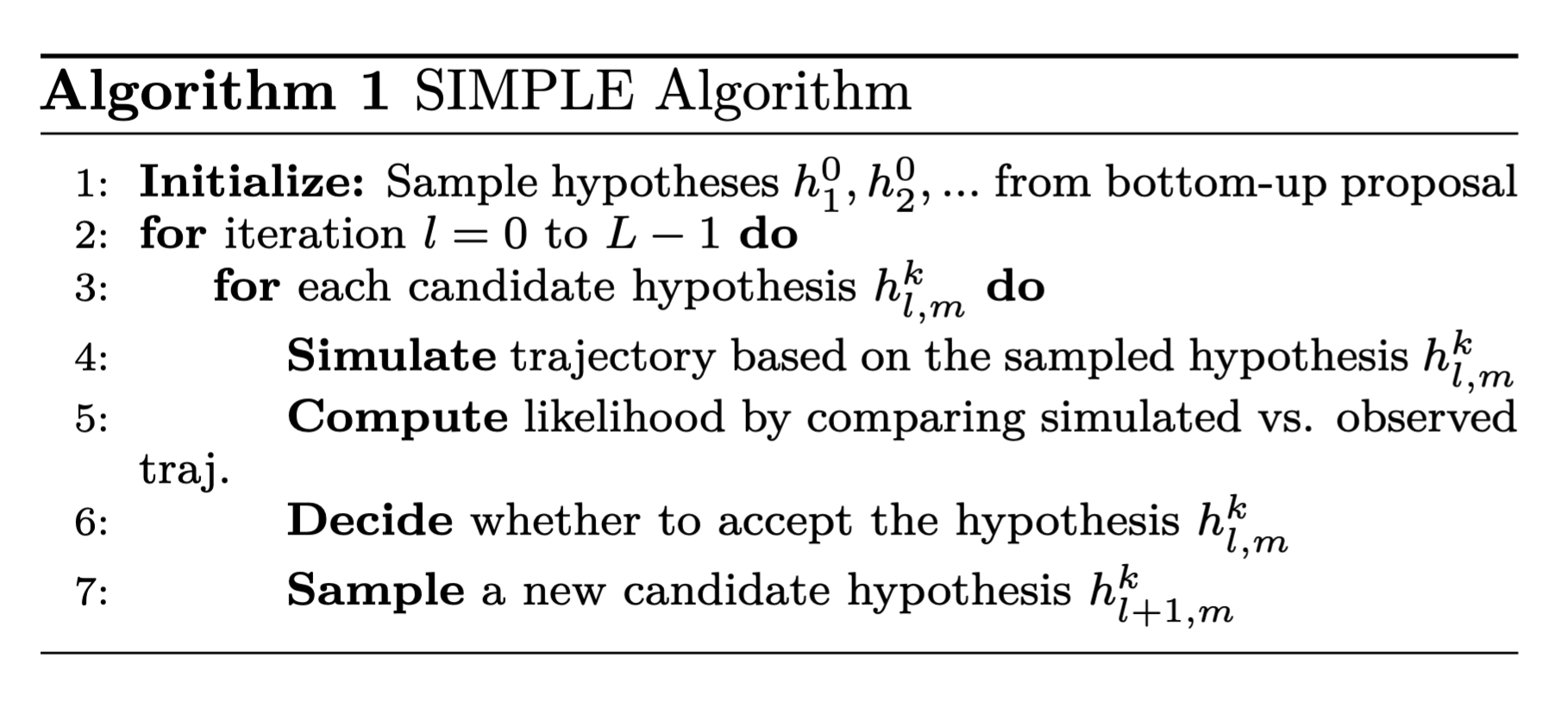}
    \end{subfigure}
    \hspace{1.5em}
    \begin{subfigure}[t]{0.22\textwidth}
        \centering
        \setlength{\unitlength}{1pt}
        \begin{picture}(0,0)
            \put(-70,-18){\large\textbf{C}}
        \end{picture}
        \includegraphics[width=\linewidth]{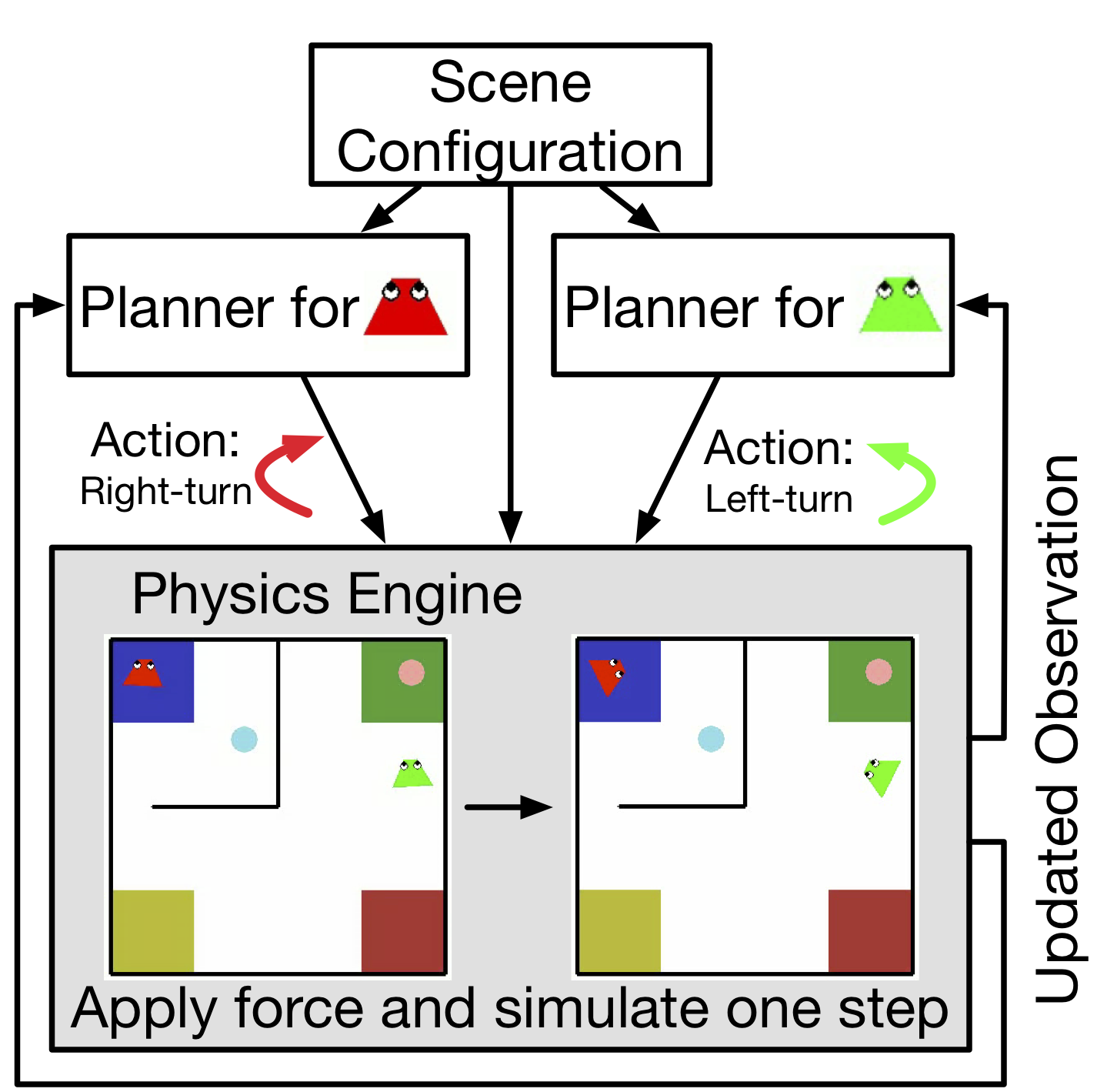}
    \end{subfigure}

    \vspace{0.5em}

    \begin{subfigure}{\textwidth}
        \centering
        \setlength{\unitlength}{1pt}
        \begin{picture}(0,0)
            \put(-20,150){\large\textbf{B}}
        \end{picture}
        \includegraphics[width=0.8\textwidth]{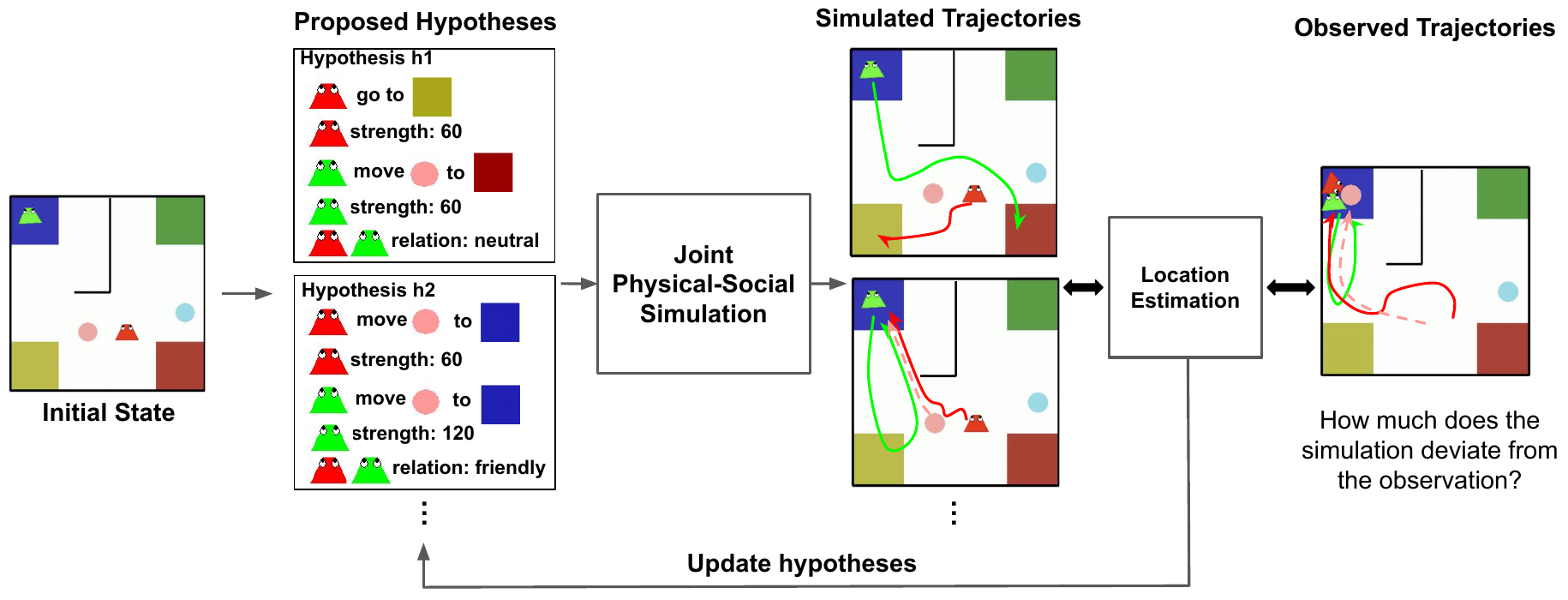}
    \end{subfigure}

    \caption{(\textbf{A}) Pseudo code for the SIMPLE model. (\textbf{B}) Illustration of the key model components of SIMPLE. (\textbf{C}) Illustration of the joint physical and social simulator in SIMPLE.}
\end{figure*}
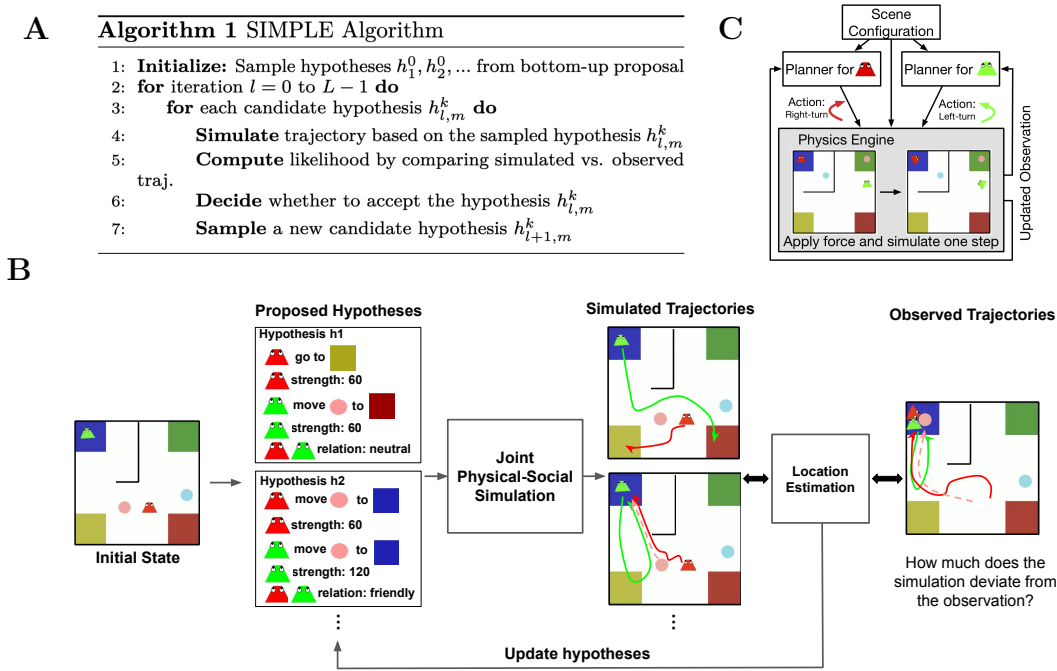

\FloatBarrier
\section{The SIMPLE Model}

We propose that human social understanding relies on the flexible integration of intuitive physics and intuitive psychology. We hypothesize that observers formulate a generative model of agent interaction and invert this model to infer rich social information from physically grounded scenes. We instantiate this approach with a computational model named SIMPLE (SIMulation, Planning, and Local Estimation). While standard inverse planning assumes agents are rational planners and infers mental states by inverting a planning model, it often abstracts away the physical environment. We hypothesize that people invert a \textit{joint} physical and social generative model. This allows for the simultaneous inference of latent variables, including agent goals, social relationships, and physical capacities (e.g., strengths). Therefore, SIMPLE extends the framework of Bayesian inverse planning by coupling computational theory of mind \citep{baker2017rational} with simulation-based physical reasoning \citep{battaglia2013simulation}.

We formalize the observer's reasoning as inference over a set of hypotheses about latent variables. Let $h = \langle g_i, g_j, \alpha_{ij}, \alpha_{ji}, f_i, f_j \rangle$ denote a specific hypothesis, where $g$ represents agent goals, $\alpha$ denotes social relationship parameters, and $f$ indicates physical strengths (i.e., the maximum amount of force an agent can exert). Let $s^{1:T}$ be the observed state sequence (specifically, the trajectories of all entities) and $\hat{s}^{1:T} = G(h)$ be the simulated trajectory given the hypothesis. The generative function $G(\cdot)$ is a closed-loop process that couples a hierarchical planner---representing intuitive psychology---with a forward physics engine---representing intuitive physics---that resolves contact dynamics and object motion.

To infer the goals and relations, we perform Bayesian inverse planning. We define the posterior probability distribution as:

\begin{equation}
\begin{array}{l}
P(h = \langle g_i, g_j, \alpha_{ij}, \alpha_{ji}, f_i, f_j \rangle | s^{1:T}) \
\propto \\
P(s^{1:T}|h)P(g_i)P(g_j)P(\alpha_{ij},\alpha_{ji})P(f_i)P(f_j),
\end{array}
\label{eq:posterior}
\end{equation}
where $P(s^{1:T}|Y) = e^{-\beta \sum_{t=1}^T ||s^t-\hat{s}^t||_2}$ is the likelihood of the observed trajectory conditioned on the hypothesis, and $\beta > 0$ is a constant coefficient. Because this inference must account for complex environment geometry, unobservable agent capacities, and diverse social interactions (e.g., helping, hindering, or neutral co-presence), the hypothesis space is high-dimensional and continuous. We therefore approximate the posterior using sampling-based probabilistic inference, outlined below.

\subsubsection{Joint Physical and Social Simulation}
We employ a joint physical and social simulator as our generative model. Unlike approaches that rely solely on symbolic planners, SIMPLE integrates a physical simulation engine with a hierarchical planner (details are provided in Appendix 3).

Given a scene configuration (the initial state) and a hypothesis $h$, the simulation proceeds in steps: (1) the hierarchical planner samples actions for all agents based on their goals and beliefs; (2) these actions are fed into the physics engine to resolve dynamics; and (3) the engine renders the next frame of the trajectory. This generates a fully physically grounded prediction $\hat{s}^{1:T}$ that accounts for interactions such as collisions or joint object manipulation. Additional information and implementation details can be found in the Appendix 4.

\subsubsection{Inference via Metropolis-Hastings with Local Estimation}

Due to the combinatorially large space of hypotheses that makes exact inference intractable, we use an efficient algorithm to approximate the posterior distribution. We first use a bottom-up approach to propose likely hypotheses (See Appendix 4B). To explore the hypothesis space and infer the posterior distribution, we utilize Markov Chain Monte Carlo (MCMC). Specifically, we employ the Metropolis-Hastings algorithm. We run multiple iterations to update the proposals. Given the $M$ proposals at iteration $l$, we simulate the trajectories, i.e., $\hat{s}^{1:T}_{l,m}$, $\forall m = 1,\cdots, M$, and compare them with the observed trajectories, $s^{1:T}$. For each proposal, we sample a time interval with a fixed length, $\Delta T$, based on the errors between the simulation and the observations, i.e., $t_{l,m} \propto e^{\eta\sum_{\tau=t_{l,m}}^{t_{l,m}+\Delta T}||\hat{s}^\tau_{l,m}-s^\tau||_2}$, where $\eta$ is the scaling factor. The intuition behind this is that local deviation is often more informative in terms of how the proposal should be updated compared to the overall deviation. After selecting a local time interval, we use the same bottom-up mechanism to again propose a new hypothesis for each particle, $h^\prime_{m}$, based only on $S^\prime=s^{t_{l,m}:t_{l,m}+\Delta T}$. We then use the Metropolis--Hastings algorithm to decide whether to accept this new proposal for the particle, where the acceptance rate is $\alpha = \min\{1,\frac{Q(h^\prime | S^\prime)P(s^{1:T}|h^\prime)}{Q(h_{l,m} | S^\prime)P(s^{1:T}|h_{l,m})}\}$, where $Q(\cdot)$ is the proposal distribution. This process allows the chain to converge to the stationary distribution corresponding to the true posterior $P(h|s^{1:T})$.

\subsubsection{Marginalization}
Finally, given the approximated joint posterior distribution, we marginalize over the hypotheses to extract the probability distributions for individual latent variables of interest, such as the probability of a specific social relation $\alpha_{ij}$ or goal $g_i$. For instance, we can compute the posterior distribution of $g_i$ as

\begin{equation}
P(g_i | s^{1:T}) = \sum_{g_j, \alpha_{ij}, \alpha_{ji}, f_i, f_j} P(g_i, g_j, \alpha_{ij}, \alpha_{ji}, f_i, f_j| s^{1:T}).
\label{eq:marginalization}
\end{equation}

\subsection{Alternative Models}
We compared SIMPLE against a range of alternative models. First, we evaluated end-to-end neural baselines, which have been popular in modeling and understanding visual scenes, including social perception. We included SocialGNN \citep{malik2023relational} and Gemini 2.5 Pro \citep{comanici2025gemini} as our neural baselines. SocialGNN is a graph neural network (GNN) that learns to predict social labels from visual scenes. Because the original SocialGNN only predicts relations, we extended the model by training it on the 500 training videos in PHASE to jointly predict the goal and relation labels. Gemini 2.5 Pro is a state-of-the-art VLM trained on large corpora of visual and language datasets. Second, we included a predicate inverse planning model, a physics-agnostic variant of SIMPLE. This model extracts high-level predicates from states and performs inverse planning to infer goals and relations. Compared to SIMPLE, this ablated model does not encode the physical dynamics necessary to reason about the fine-grained motions of agents and objects in the scenes. See Appendix 4 for implementation details.

\begin{figure*}[ht!]
    \centering
    \includegraphics[width=0.95\textwidth]{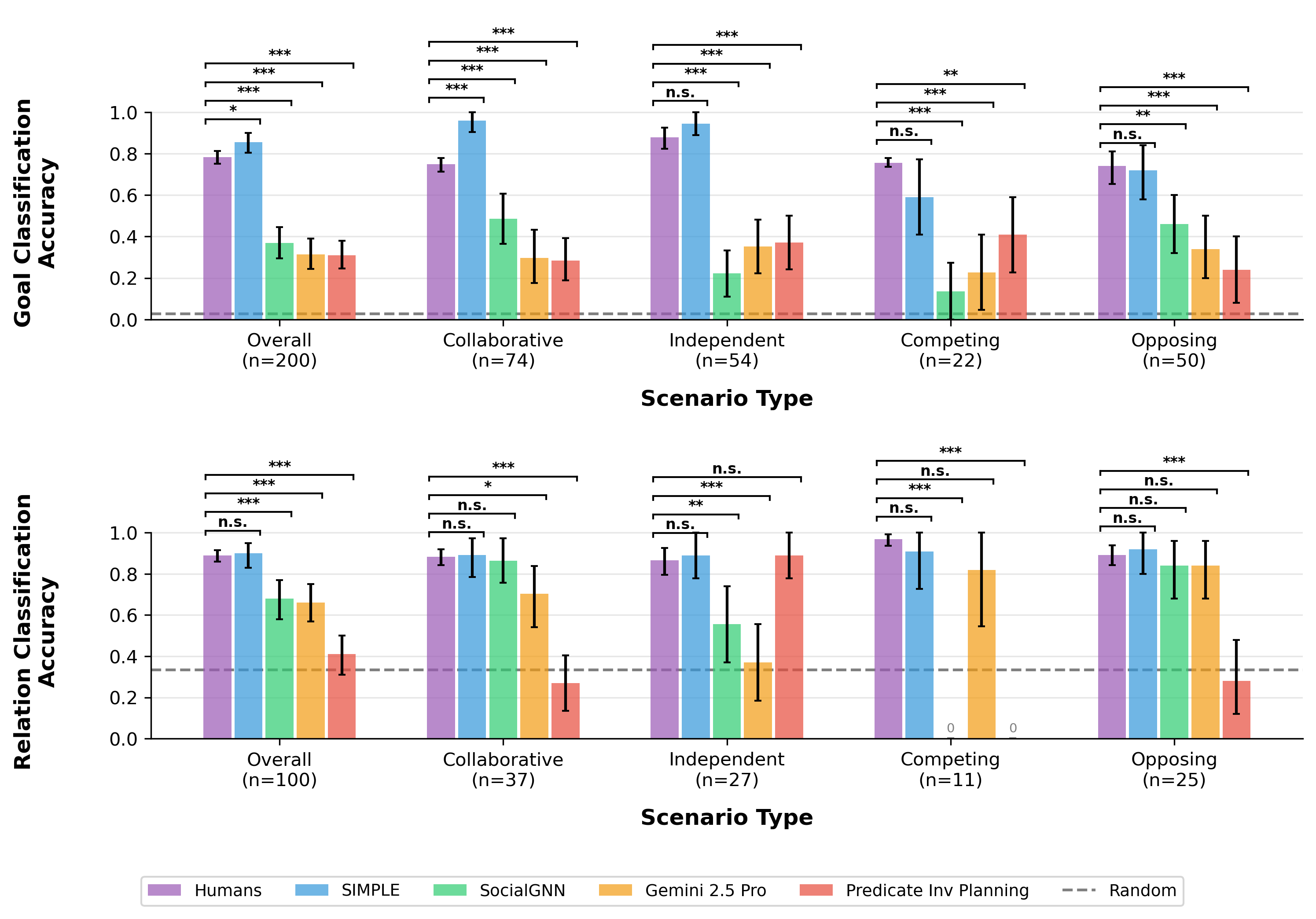}
    \caption{Accuracy results for goal classification task across all 100 scenarios and grouped by 4 distinct scenario goal types. The number of goal judgments is shown in the bracket. Humans and models are evaluated on 100 test videos, each with two goal classification tasks and one relation classification task. Error bars show 95\% confidence interval from 1000 bootstrapped samples.}
    \label{fig:accuracy}
\end{figure*}

\begin{figure}[h!]
    \centering
    \includegraphics[width=0.75\linewidth]{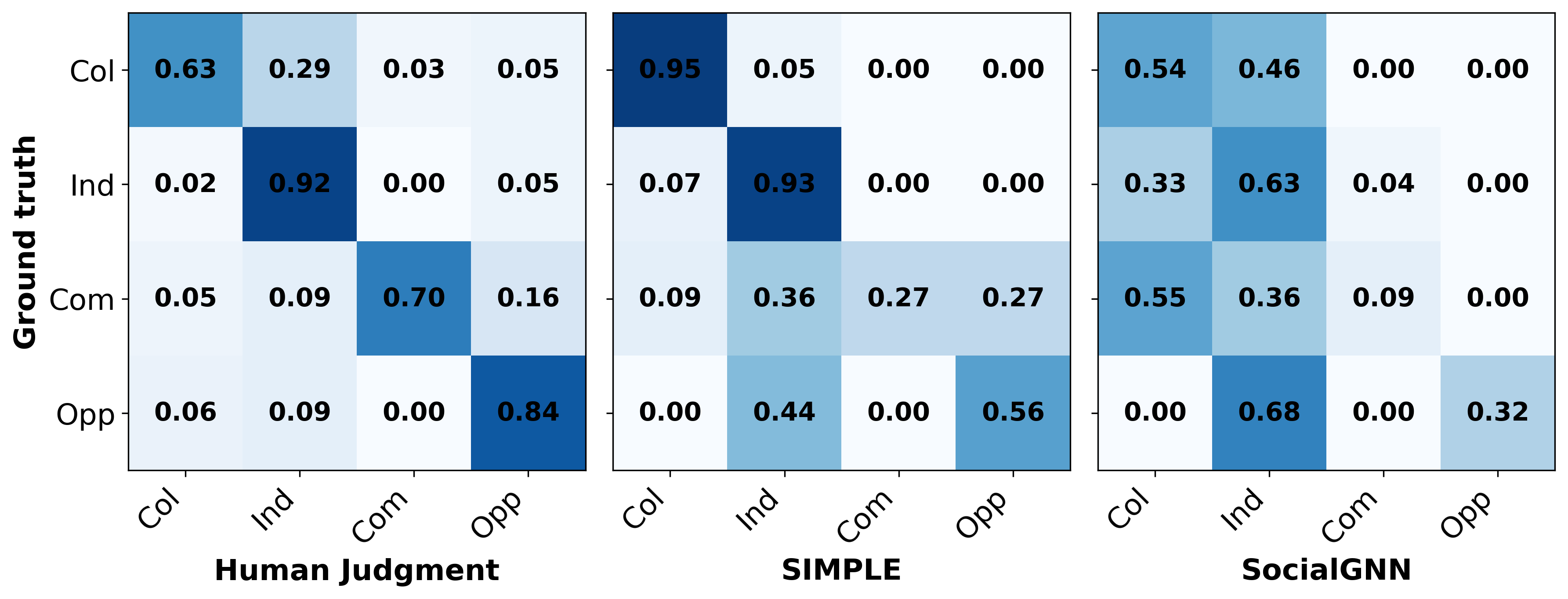}
    \caption{Confusion matrix of SIMPLE and SocialGNN on goal classification tasks. (Col = Collaborative, Ind = Independent, Com = Competing, Opp = Opposite)}
    \label{fig:confusion}
\end{figure}

\begin{figure*}[ht!]
\centering
    \includegraphics[width=\linewidth]{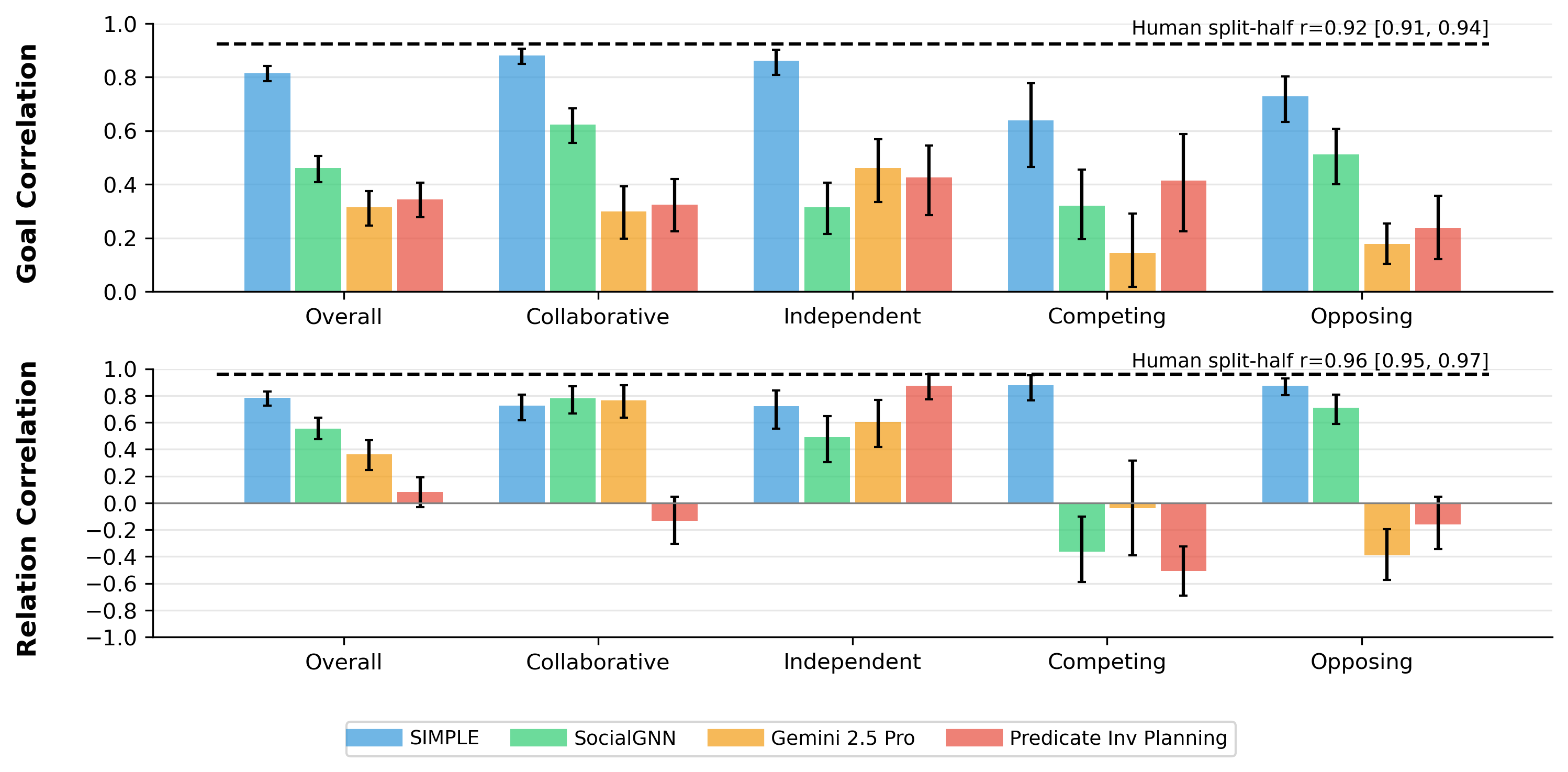}
    \caption{Correlation between model and human judgments tasks. }
    \label{fig:correlation_bar}
\end{figure*}

\begin{figure}[t!]
    \centering
    \includegraphics[width=0.75\linewidth]{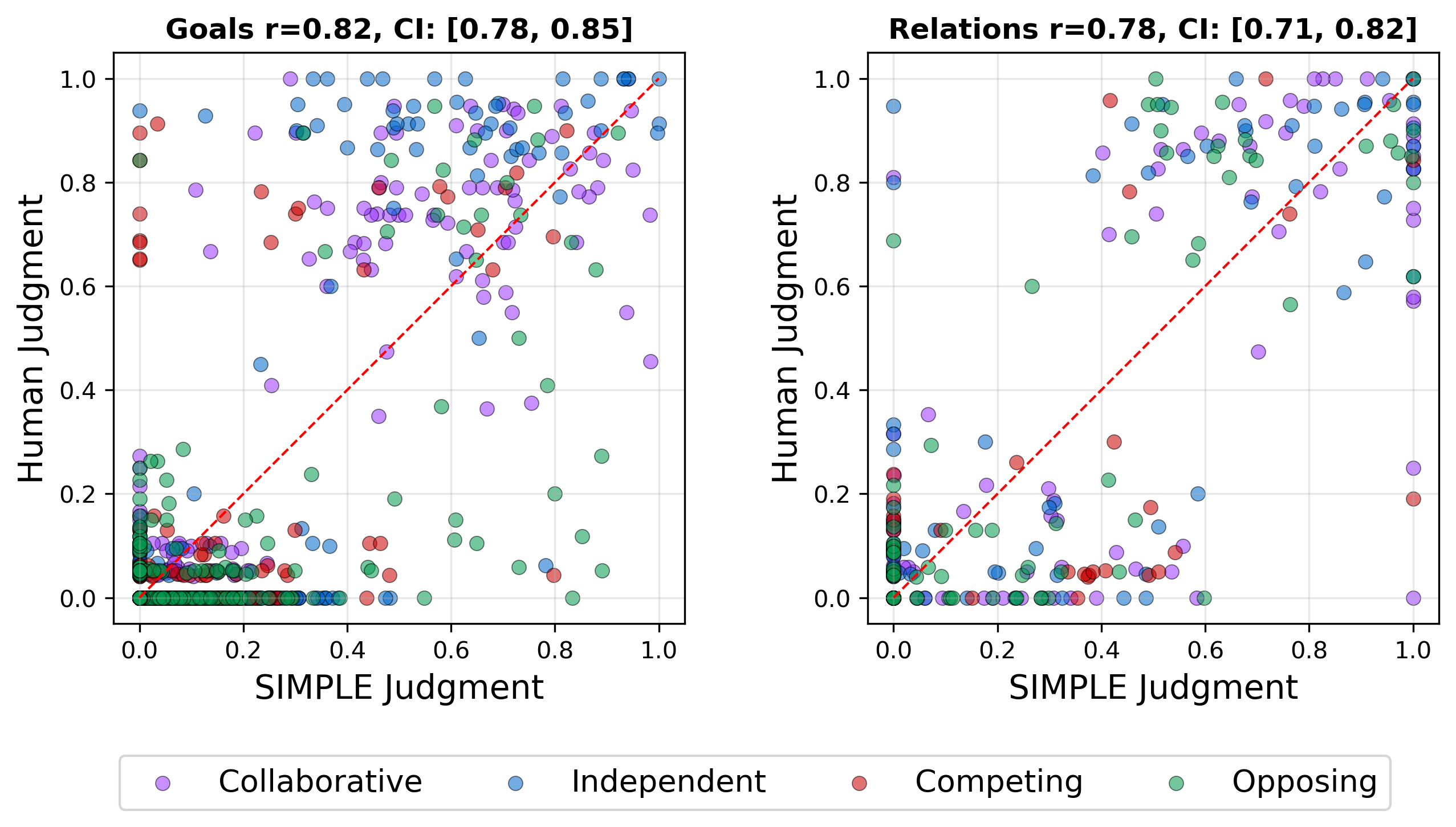}
    \caption{Scatterplots comparing human judgment against SIMPLE model judgment on goals (left) and relations (right). The colors indicate the scenario types.}
    \label{fig:correlation_scatter}
\end{figure}

\begin{figure*}[t!]
    \centering
    \includegraphics[width=\linewidth]{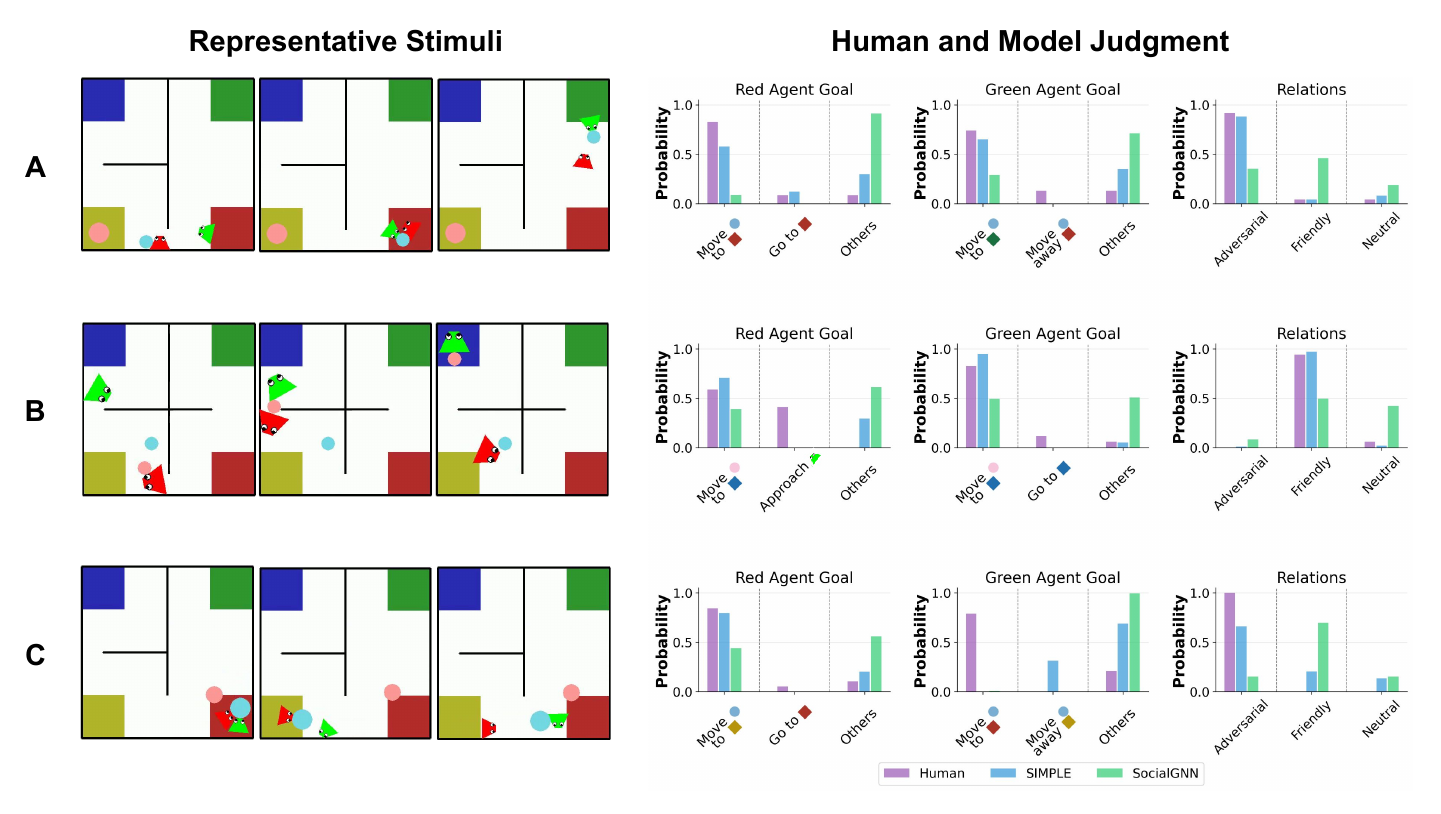}
    \caption{Qualitative examples showing human and model judgments on PHASE animations. Examples \textbf{A} and \textbf{B} show a competing and collaborative scenario, respectively. Across both scenarios, SIMPLE aligns with human distribution on goal and relation classifications, whereas SocialGNN fits less well. Example \textbf{C} shows a competitive scenario where SIMPLE misaligns with humans. In this example, the two agents want to push the blue circle to different landmarks. The green agent eventually steals the blue circle from the red agent. SIMPLE was able to correctly recognize the goal of the red agent, but misclassified the green agent's goal as hindering the red agent (move the blue circle away from the yellow landmark) and other independent goals (e.g., moving to the red landmark).}
    \label{fig:qualitative}
\end{figure*}

\section{Results}
We compared human participants, SIMPLE, and alternative computational models on the same 100 test videos and compared their judgments with human judgments. Among all models, SIMPLE not only achieved the highest accuracy in goal and relation accuracy, but also best captures human judgments. We summarize the key results below.

\subsection{Human results}
We recruited 200 participants (Mean age = 39.95, 89 Female, Male 105, 6 Non-binary) to judge the agents' goals and relations for the 100 test videos (each video is rated by 20 people). We found that participants were highly consistent in their judgments. To quantify this consistency, we computed the split-half correlation among the human participants (the average correlation between the average responses of 1000 random splits). The agreement was high for both tasks ($r=0.92, \text{CI} = [0.91, 0.94]$ for goals; $r=0.96, \text{CI}=[0.95, 0.97]$ for relations), indicating that participants largely shared the same interpretations of the PHASE animations. We then computed classification accuracy against ground truth. As shown in Figure~\ref{fig:accuracy}, participants achieved high classification accuracy on average, with an accuracy of 0.89 on relation classification and 0.78 on goal classification. We provide more details of correlation analysis in Appendix 5C.

\subsection{Comparing model and human accuracy on classification tasks}
The model classification accuracy is illustrated in Figure~\ref{fig:accuracy}. Among all models, SIMPLE achieves the highest overall accuracy in both goal and relation classification tasks.

We then break down the classification accuracy by scenario types. Overall, the SIMPLE model achieved high classification accuracy in goals and relations across scenario types. Notably, there was no statistical significance between human accuracy and SIMPLE accuracy in three of the four scenario types for goal classification and all scenario types for relation classification. On the other hand, alternative models scored much lower on goal classification tasks compared to relation classification.

Among the goal classification tasks, competing scenarios appear to be the most difficult for SIMPLE (mean accuracy = 0.59, 95\% CI = [0.41, 0.77]), whereas the human average is 0.76, 95\% CI = [0.74,0.78]. We show the confusion matrix for humans, SIMPLE, and SocialGNN in Figure~\ref{fig:confusion}, which indicates that while humans accurately recognized goals in competing scenarios most of the time, SIMPLE often classified goals in competing scenarios as opposing or independent. On the other hand, SocialGNN frequently confused competing goals as collaborative instead. This explains why SIMPLE still achieved relatively high relation classification accuracy (mean = 0.91, 95\% CI = [0.72, 1.0]) while SocialGNN had a relation classification accuracy of 0.

Qualitative analysis indicates that competitive scenarios often exhibit a ``tug-of-war'' dynamic, in which two agents remain attached to the same object while exerting opposing forces. This pattern is illustrated in the diagnostic example shown in Figure~\ref{fig:qualitative}\textbf{A}. For models that primarily rely on motion cues, such as SocialGNN, the correlated movement of two agents alongside a shared object can resemble cooperative behavior---e.g., agents jointly pushing an object toward a common landmark. As a result, the interaction may be misclassified as friendly. In contrast, SIMPLE employs an analysis-by-synthesis approach, enabling it to detect that the irregular and conflicting motion patterns are inconsistent with a shared goal of transporting the object toward the same destination. Although SIMPLE may still struggle to fully interpret the scene or assign the correct interaction label, it more effectively captures the underlying competitive dynamics. This limitation is further illustrated in Figure~\ref{fig:qualitative}\textbf{C}, where the model incorrectly attributed opposing goals within a competing scenario.

The Predicate Inverse Planning model performed well at classifying neutral relations in Independent scenarios. This suggests that observers may not always require detailed physical simulation to interpret social relations in relatively simple contexts. For example, if Agent A moves toward a red circle while Agent B moves toward a blue one, symbolic predicates alone may be sufficient to infer that their relationship is neutral. However, the same model struggled with goal classification in these scenarios, achieving a mean accuracy of 0.35 (95\% CI = [0.22, 0.48]). This dissociation indicates that while broad relational categories can sometimes be derived from symbolic cues, inferring specific physical goals---such as whether Agent A is capable of pushing the circle toward a landmark given its strength---demands a more fine-grained understanding of the underlying physical constraints. Such inferences are better captured by a physically grounded inverse planning framework like SIMPLE.

\subsection{Explaining human graded judgments}
Human social perception and judgment are rarely unanimous, and it involves graded degrees of belief and uncertainty within and across people. To evaluate whether models capture this nuance, we compared the models' posterior probability distributions against the empirical distribution of human responses on both goal and relation classification tasks.

Figure~\ref{fig:correlation_bar} shows the Pearson correlation between model predictions and human judgments on both goal and relation classification tasks. SIMPLE achieves a strong correlation against human distribution ($r = 0.82$, 95\% CI = [0.78, 0.85] for goals; $r = 0.78$, 95\% CI = [0.71, 0.82] for relations). In contrast, baseline correlations are significantly lower ($r < 0.6$).

Figure~\ref{fig:correlation_scatter} presents scatterplots of model judgments against human judgments, revealing shared patterns of uncertainty across the two. A comparison of the plots shows that both human participants and SIMPLE exhibited greater uncertainty in goal judgments than in relation judgments, as reflected by the concentration of data points in the 0.6--0.8 range. This is because the goal space is substantially larger and more complex than the relation space. As a result, identifying an agent's specific goal from limited observational data is inherently more challenging than assigning a broader relational label.

\section{Discussion}

We presented the PHASE dataset, consisting of synthetic animations in the style of Heider and Simmel movie, to study how people perceive social scenes in physically grounded scenarios. We also proposed a computational model, SIMPLE, which integrates a forward physics engine into a Bayesian inverse planning framework. By doing so, it achieved high accuracy and correlation with human judgments across diverse social interaction scenarios and explained human judgments better than baselines, including model-free, feedforward models---such as graph neural networks and large vision-language models---and a purely symbolic-predicate based inverse planning model.

The comparison between feedforward baselines and our SIMPLE model highlights the need for an analysis-by-synthesis type of social inference for reverse engineering human-like, physically grounded social perception. Feedforward recognition models either learn predictive visual representations from provided training videos of agent social interactions in PHASE environments (e.g., SocialGNN) or visual reasoning capacities emerged from internet-scale pretraining (e.g., Gemini 2.5 Pro). These models treat social perception as a strictly feedforward process, mapping visual patterns to social judgments. Prior work has demonstrated that discriminative visual patterns can be predictive of basic aspects of agent behavior, such as animacy \citep{hovaidi2018neural,shu2021unified}, intentionality \citep{epstein2020oops}, interactivity \citep{shu2018perception}, shared attention \citep{fan2018inferring}, and relationships \citep{hovaidi2018neural,malik2023relational}. However, unlike people or SIMPLE, these feedforward models require a large amount of training data; struggle with generalization to unseen social interactions or trivial alterations (such as different physical environments); and cannot robustly make inferences of agents' mental states, such as goals as demonstrated in our experiment, beliefs \citep{ullman2023large,shapira2023clever,mmtomqa}, and desires \citep{shu2021agent}. They also do not possess the sophisticated physical reasoning abilities required to understand the social behavior grounded in complex physical environments. This is particularly evident in the low recognition accuracy of SocialGNN and Gemini 2.5 Pro in competing scenarios where agents engage in much more complex and diverse physical interactions with one another and the objects compared to other scenarios. In particular, similar to our experiment, prior work has also observed a similar lack of physical common sense in VLMs \citep{gao2025vision}, which could partially explain Gemini 2.5 Pro's poor understanding of PHASE videos. By inverting the generative process of how agents plan to act under the physical dynamics of a given environment, SIMPLE produced much more human-aligned social judgment without the need for model training. SIMPLE's core ability to simulate physically grounded social behavior matched with prior findings on how people rely on physical concepts---such as efficient actions \citep{liu2017ten,jara2020naive,shu2021agent} and danger \citep{liu2022dangerous}---and extended the prior work from single-agent behavior to multi-agent interactions.

SIMPLE also differs from prior work on inverse planning, which largely relies on symbolic representations of agent behavior and environments and focuses on inverting only the action planning process. In particular, SIMPLE provides a computational account of the integration of intuitive physics and intuitive psychology. First, as agents' actions are constrained by physics. Their planning processes naturally consider their own physical capacities and constraints (such as their strengths and sizes). They also plan their actions to actively alter the dynamics of the physical interactions, including agent-environment, agent-object, and agent-agent interactions. Second, the simulation of the physical consequences of the agents' actions gives rise to a more fine-grained estimation of behavior likelihood given hypotheses of the social interpretations of the observed motion trajectories. While feedforward models may learn predictive visual cues from large-scale training, analysis-by-synthesis style inference instantiated in SIMPLE provides a more principled and generalizable way to interpret the social meanings behind complex motion trajectories. Indeed, people can flexibly understand agent behavior in different kinds of visual stimuli, even abstract animations such as the Heider and Simmel movie or the PHASE videos, without the need for any training data. SIMPLE, by inverting a joint physical and social generative model as an approximated human mental simulation, can capture the strong generalizability of physically grounded human social perception.

The joint physical and social simulation in the SIMPLE model echoes the core knowledge system in human cognition \citep{spelke2007core}. The physical simulation, by encoding physical dynamics (such as how entities move and respond to forces) and constraints (such as the sizes of the entities and the gaps between walls), represents concepts of objects and geometry. Coupled with concepts of agents' rational behavior, the joint physical and social simulation endows our SIMPLE model with crucial foundations for commonsense physical and social scene understanding.

\subsection{Limitations}
Our approach has several limitations. First, as noted in qualitative analyses, physical ambiguity can arise during sustained contact. In hindering interactions, once agents make contact, the specific target goal often becomes ambiguous in the likelihood function (e.g., when being blocked by another agent, it is often unclear which landmark the agent is trying to reach). While the trajectory segment \textit{prior} to contact often reveals the intention, a purely instantaneous frame-by-frame likelihood can struggle without longer temporal integration.

Second, the computational cost of SIMPLE is high. Bayesian inverse planning requires many simulations per inference iteration for multiple iterations. While this captures human social reasoning, it likely does not reflect the mechanisms of rapid, real-time social perception that people often perform effortlessly. Future work can improve the efficiency of the SIMPLE model through performance engineering, like recent approaches in Sequential Monte Carlo methods for rapid online social inference \citep{zhi2020online}.

Third, human social perception can be resource rational. The brain likely employs a noisy physics engine to perform coarse physical simulation \citep{battaglia2013simulation,ullman2017mind,schwettmann2018evidence,pramod2020evidence}. In many cases, people may resort to simple heuristics that do not warrant physical simulations at all. Such fast recognition mechanisms could be accounted for by amortized inference \citep{gershman2014amortized,jha2024neural,jha2025modeling}, in which a feedforward model can be trained to match its predictions with posterior distributions produced from explicit probabilistic inference. We intend to investigate how to amortize the joint physical and social inference in SIMPLE via neural model training in future work. We also plan to probe when and what kind of physical simulation is needed.

Lastly, our domain is limited to 2D geometric shapes. While this follows the rich tradition of the Heider and Simmel movie, real-world social perception involves articulated 3D bodies, gaze, and fine-grained motor control. The physics of 2D discs is a simplified proxy for the complex biomechanics of human movement. There have been recent works that leverage pretrained LLMs and VLMs to achieve open-ended inverse planning \citep{ying2023neuro,zhi2024pragmatic, mumatom,thoughttracing,autotom, ying2024pragmatic,jha2025modeling}. While these models can conduct both efficient and robust mental state inferences given complex, real-world stimuli, they rely on symbolic representations of states and actions. It remains unclear whether and how these types of hybrid models can generalize to understanding the social and physical dynamics of continuous motions.

\section{Materials and Methods}

The PHASE dataset, all the qualitative examples featured in this paper, as well as the code for the SIMPLE model, can be accessed via \href{https://osf.io/fkp5m/}{https://osf.io/fkp5m/}.

All human experiments are conducted through a customized online interface (\href{https://phase-interface.web.app/}{https://phase-interface.web.app/}). All participants provided informed consent, and the study was approved by the MIT Institutional Review Board.

\bibliographystyle{unsrtnat}

\newpage
\appendix
\renewcommand{\thesection}{Appendix \Alph{section}}
\renewcommand{\thesubsection}{\Alph{section}.\arabic{subsection}}
\renewcommand{\thefigure}{\Alph{section}\arabic{figure}}
\renewcommand{\thetable}{\Alph{section}\arabic{table}}
\renewcommand{\theequation}{\Alph{section}\arabic{equation}}
\renewcommand{\thealgorithm}{\Alph{section}\arabic{algorithm}}

\section{Human Evaluation of the PHASE Dataset}
\setcounter{figure}{0}
\setcounter{table}{0}
\setcounter{equation}{0}

\subsection{Human video label tagging for PHASE videos}
The 23 labels used in this experiment are: not interacting, interacting unintentionally, chasing, running away, stalking, approaching, avoiding, meeting, gathering together, guiding, following the lead (of another agent), playing a game of tag, blocking, fighting, competing, stealing, protecting an object, attacking, hindering, bullying, playing tug of war, helping, collaborating.

\subsection{Comparing PHASE and human-generated videos}

We recruited 3 human participants to create videos in the same 2D environment as PHASE. The human video generation procedure is similar to the setup in PHASE, except that actions are obtained from human user input. In each game, there are two players, one for controlling each agent. The players each view the environment and control their agent from separate screens. Players can use the following actions by pressing keys on their keyboards: 4 directions (forward, backward, right, left), turning right or left, and grabbing or letting go of an object. We reset the velocity of each agent to 0 after each step to make it easier for players to control the agents. Before each session, the players were shown a tutorial on how the agents work (partial observability and the controls). They were given an opportunity to play freely in the game environment to get familiar with the controls. At the beginning of a session, they were told the goals assigned to both players (so they knew each other's goals) and asked to start playing the game to achieve the assigned goals. Each session ended either when the goals of both players were achieved or when the time limit was reached.

We then recruited 186 participants on Mechanical Turk and randomly divided them into two groups. One group watched the human-controlled videos, and the other watched matching videos from PHASE. For each video, participants were asked to judge the goals and relations of the agents, and rate how likely humans were to behave similarly to these agents under the same goals and relations (on a scale of 1 to 5). We compare the mode human response against the ground truth. In both groups, the mode of participants' responses produces a high accuracy for goal and relation recognition ($0.965$ and $0.92$ for goal and relation recognition on the human-controlled videos, and $0.97$ and $0.99$ for goal and relation recognition on the PHASE videos). The averaged human-likelihood rating for the human-controlled videos is $4.06$ ($\sigma=0.36$); and for PHASE, it is $3.98$ ($\sigma=0.42$). This suggests that to the participants, (i) the PHASE videos and the human-controlled videos exhibit similar social events in terms of goals and relationships, even though they have different motion trajectories, and (ii) the agent behaviors in these two types of videos all have similar degrees of human-likelihood.

\section{Main Experiment on Collecting Human Goal and Relation Judgments on Videos}
\setcounter{figure}{0}
\setcounter{table}{0}
\setcounter{equation}{0}

\begin{figure}[t!]
    \centering
    \includegraphics[width=0.7\linewidth]{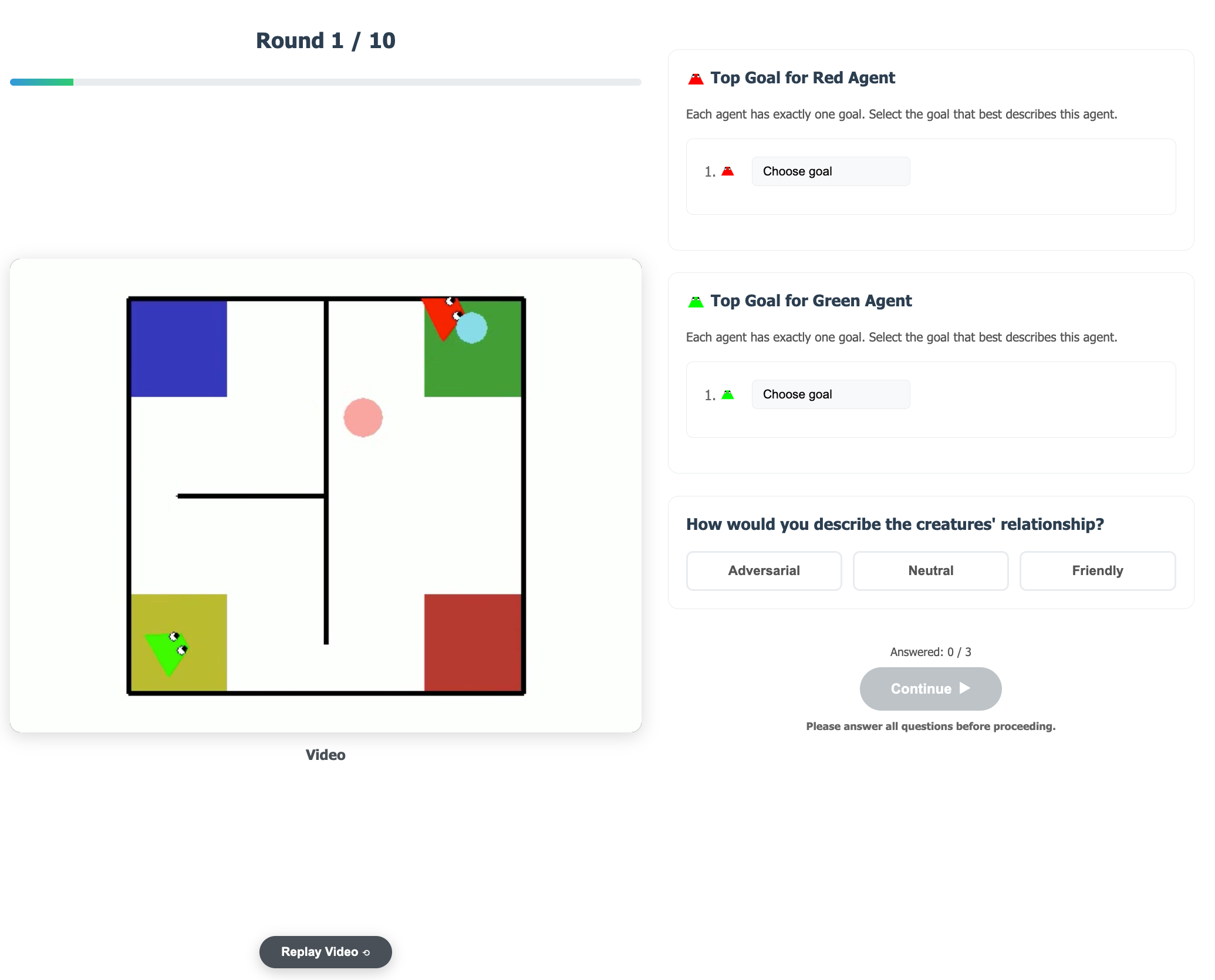}
    \caption{Interface used to collect human judgments of agent goals and social relations from PHASE videos. Participants selected the most likely goal for each agent and indicated the agents' relationship after viewing each video.}
    \label{fig:interface}
\end{figure}

Human goal and relation judgments were collected using a custom web-based annotation interface applied to videos from the PHASE dataset (Fig.~\ref{fig:interface}). Each participant completed 10 trials. On each trial, participants viewed a PHASE interaction video and selected the most likely goal for each agent from a predefined set of candidate goals. Participants also indicated the perceived social relationship between the agents (adversarial, neutral, or friendly). The interface displayed the video on the left and the response panel on the right, allowing participants to inspect the interaction while making selections. A progress indicator provided feedback on task completion. Prior to the main task, participants completed a brief tutorial and a comprehension quiz designed to ensure understanding of the agents' dynamics and the response procedure. Selections and response timestamps were recorded automatically under anonymized participant identifiers and stored in a cloud-hosted database for analysis.

\section{Details of Joint Physical-Social Simulation}
\setcounter{figure}{0}
\setcounter{table}{0}
\setcounter{equation}{0}
\setcounter{algorithm}{0}

\begin{algorithm}[t!]
\caption{Joint Physical-Social Simulation}\label{alg:simulation}
\begin{algorithmic}[1]
\STATE \textbf{Input:} $g_1, g_2, \alpha_{12}, \alpha_{21}, f_1, f_2$, and initial state $s^1$
\STATE \textbf{Output:} Abstract social event $s^{1:T}$
\FOR{agent $i = 1, \dots, 2$}
    \STATE Initialize belief particles $\{b^0_{i,k}\}_{k=1}^K$
\ENDFOR
\FOR{time steps $t = 1, \dots, T$}
    \FOR{agent $i = 1, \dots, 2$}
        \STATE Update observation $o^{t}_i$
        \STATE Update belief particles $\{b^{t}_{i,k}\}_{k=1}^K$ based on $o_i^t$
        \STATE Set the other agent $j \leftarrow \{1,2\} \setminus \{i\}$
        \FOR{each particle $k = 1, \dots, K$}
            \STATE Get subgoal $h^t_{i,k} \leftarrow \text{HP}(g_i, g_j, \alpha_{ij}, b^t_{i,k})$
        \ENDFOR
        \FOR{subgoal $h \in \mathcal{H}$}
            \STATE Estimate value:
            \STATE $V(\mathcal{B}_i^t, h, g_i, g_j, \alpha_{ij}) = \frac{1}{K}\sum_{k=1}^K \mathbbm{1}(h=h_{i,k}^t) - \frac{\lambda}{\sum_{k=1}^K \mathbbm{1}(h=h_{i,k}^t)}\sum_{k=1}^K \mathbbm{1}(h=h_{i,k}^t)\hat{C}(b_{i,k}^t, s_g)$
        \ENDFOR
        \STATE Select subgoal $h_{i,*}^t = \text{argmax}_{h} V(\mathcal{B}_i^t, h, g_i, g_j, \alpha_{ij})$
        \STATE Get belief particles $\tilde{\mathcal{B}}_i^t$ that correspond to $h_{i,*}^t$
        \STATE Get action $a_i^t \leftarrow \text{LP}(\tilde{\mathcal{B}}_i^t, h_{i,*}^t)$
    \ENDFOR
    \STATE Update state $s^{t+1} \leftarrow \mathcal{T}(s^t, \{a_i^t\}_{i=1}^2, \{f_i\}_{i=1}^2)$
\ENDFOR
\end{algorithmic}
\end{algorithm}

Our joint physical-social simulation is described in Algorithm~\ref{alg:simulation}, which includes a physical simulation $\mathcal{T}$, and a hierarchical planner which consists of a high-level planner (HP) and a low-level planner (LP). Given the scene configuration, the simulation updates the belief particles based on new observations, uses the hierarchical planner to sample actions for all agents based on the updated particles, feeds the actions to the physics engine to simulate one step, and renders 5 frames of video based on the simulated physical states. The final video has a frame rate of 20 FPS. We discuss more implementation details as follows.

\subsection{Predicates, Symbolic States, Goals, and Subgoals}

\begin{table*}[h!]
    \centering
    \begin{tabular}{r|l}
    Predicate  &  Definition \\ \hline
       \textsc{On}(\emph{agent/object}, \emph{landmark})  &  An entity is on a landmark\\
    \textsc{Touch}(\emph{agent}, \emph{agent/object)} & An agent touches another entity\\
     \textsc{Attach}(\emph{agent}, \emph{object}) & An object is attached to an agent's body\\
     \textsc{Close}(\emph{agent/object}, \emph{agent/object/landmark}) & An entity is within a certain distance away from another entity or a landmark
    \end{tabular}
    \caption{Predicates and their definitions. Note that we also consider their negations, which are not shown in the table for brevity.}
    \label{tab:predicates}
\end{table*}

In our simulation, we define a set of predicates as summarized in Table~\ref{tab:predicates}. These predicates and their negations are used to (i) convert a physical state into a symbolic state, and also (ii) become a subgoal space that our hierarchical planner considers for the high-level plans.

Furthermore, the final goal states for physical goals and social goals of agents are also represented by a subset of these predicates, i.e., \textsc{On}(\emph{agent/object}, \emph{landmark}), \textsc{Touch}(\emph{agent}, \emph{agent)}, and their negations.

\begin{figure*}[t!]
\centering
\includegraphics[width=0.85\textwidth]{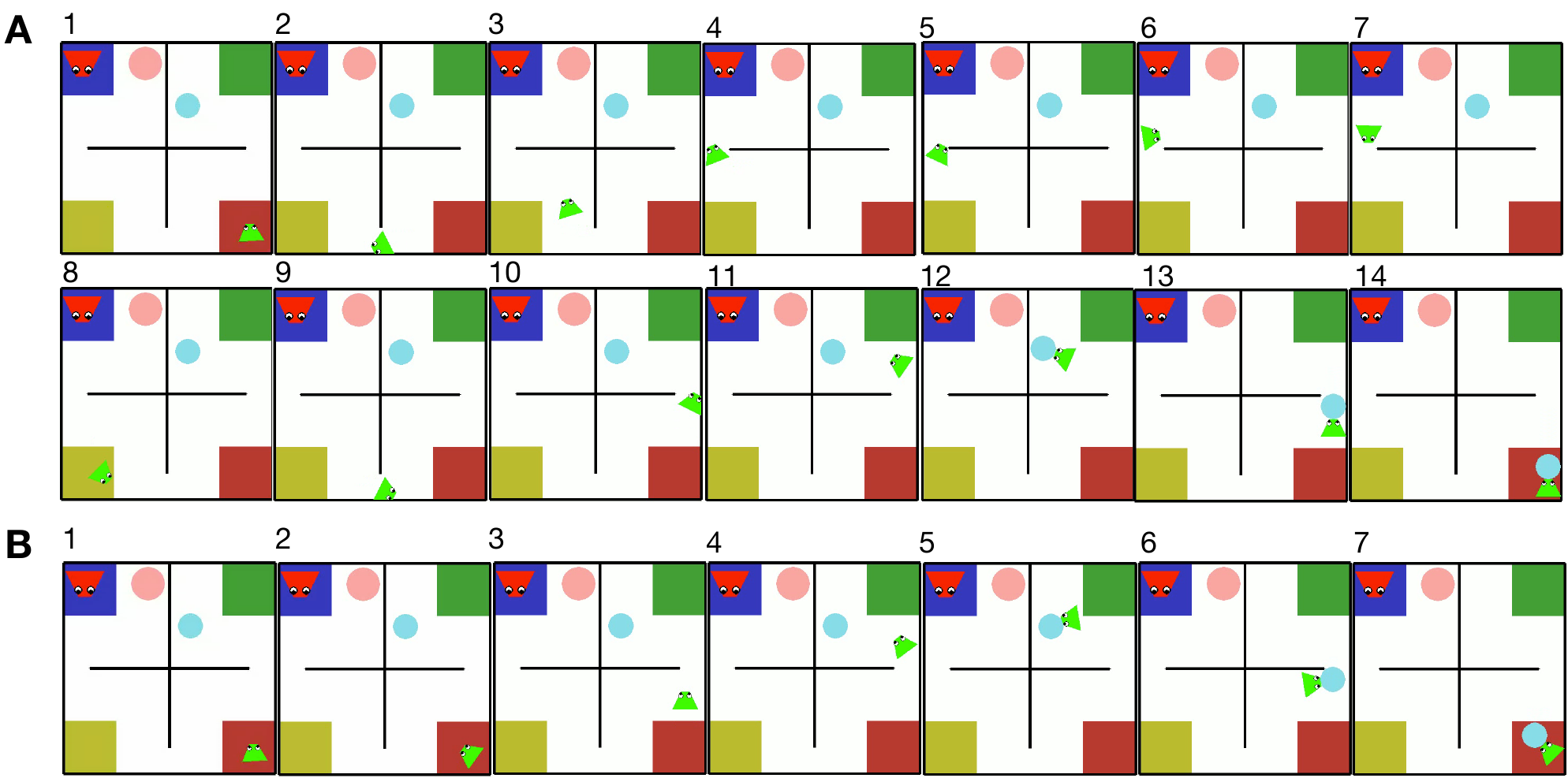}
\caption{Illustration of the effect of the estimated value function for the high-level planner. The numbers indicate the temporal order of the frames. In both sequences, the green agent's goal is to move the blue object to the red landmark in the bottom-right corner. Since it does not see the blue object initially, it needs to first find the object. ({\textbf{A}}) The sequence when $\lambda=0$. Since there is more unseen space in the left part of the environment, it is more likely that the blue object is in the left part. So the green agent first searches the left part when not considering the cost of doing so.  (\textbf{B}) The sequence when $\lambda=0.05$. When considering the cost, it is worthwhile for the green agent to search the nearby region first. The chance of finding the blue object there is slightly lower than in the left region, but the resulting cost is considerably lower. In particular, it first looks around (frame \#2) and then proceeds to search the upper-right part (frame \#3 and \#4). This comparison demonstrates that an appropriate $\lambda$ could give us more natural agent behaviors under partial observability.}
\label{fig:value_example}
\end{figure*}

\subsection{Hierarchical Planner}

For the \textbf{high-level} planner, we use A$^*$ to search for a plan of subgoals for $N=2$ agents based on $K=50$ belief particles.

To ensure a subgoal selection for simulating natural agent behavior without expensive computation, we design a heuristics-based value estimation $V(\mathcal{B}_i^t,h,g_i,g_j,\alpha_{ij})$ for each subgoal as shown in Algorithm~\ref{alg:simulation}. This value function favors subgoals that are more likely to be the best subgoal in the true state (i.e., high frequency subgoals generated by all belief particles) and have lower cost (i.e., $\hat{C}$ estimated by the distance from the current state to the final goal state according to a given belief particle). By changing the weight $\lambda$, we are able to alter the agent's behavior. Figure~\ref{fig:value_example} demonstrates an example of how $\lambda$ affects the agent's behavior. In practice, we find $\lambda=0.05$ offers a good balance and can consistently generate natural behaviors.

For the \textbf{low-level} action planner, we use POMCP \citep{silver2010monte} with
1000 simulations and 10 rollout steps. For exploration in POMCP, we adopt a variant of the PUCT algorithm introduced in Silver et al.\ \citep{silver2018general}, where we use $c_\text{init}=1.25$ and $c_\text{base}=1000$.

\subsection{Belief Representation and Update}
Each agent's belief is represented by $K=50$ particles in the simulation. Each particle represents a possible world state that is consistent with the observations. The state in a particle includes the environment layout, and physical properties of each entity --- shape, size, center position, orientation of the body, linear and angular velocity, and attached entities.

Each particle is updated with the ground truth properties of \textit{observed} entities: the agent itself, other entities in its field of view (approximated by $1\times 1$ grid cells on the map) or entities in contact with the agent. Entities that are in contact with observed entities are also defined as observed.\footnote{This is to ensure that the agent knows (i) whether there is \textit{any} other agent grabbing the same object it is currently grabbing, and (ii) whether an observed agent is grabbing \textit{any} object.} Contact occurs when entities are attached or collide, and is signaled by agents' touch sensory.

Unobserved entity properties differ between particles. We start by randomly sampling possible initial positions from the 2D environment and setting other properties (orientation and velocity) to 0. To update a belief particle from $t$ to $t+1$, we first apply the physics engine to simulate one step, where we assume constant motion for entities. Then we check the consistency between the simulated state at $t+1$ and the actual observation at $t+1$. For entities that contradict the observation, we resample their positions and orientations. We then repeat the consistency check and resampling until there is no conflict.

\begin{figure}[t!]
\centering
\includegraphics[width=0.50\textwidth]{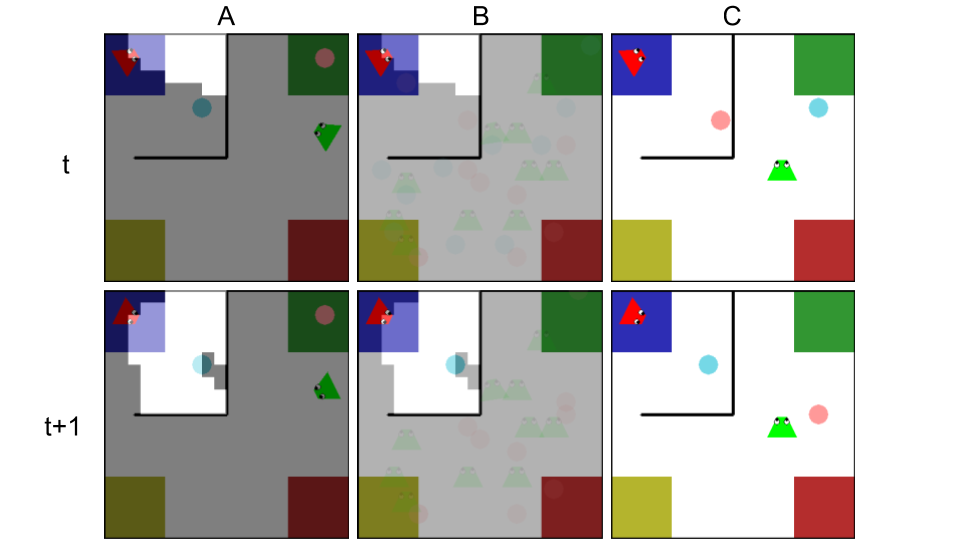}
\caption{Illustration of how the red agent updates its belief using $K=10$ particles. (\textbf{A}) True states $s^t$ (top) and $s^{t+1}$ (bottom). The bright pixels indicate the red agent's field of view.
(\textbf{B}) $\{b^t_{\text{red},k}\}^K_{k=1}$ (top) and $\{b^{t+1}_{\text{red},k}\}^K_{k=1}$ (bottom). The states in all the particles are visualized together. At step $t+1$ the red agent observes the blue object via its field of view. All particles are then updated accordingly with the ground truth properties of the blue object, and the inconsistent belief states are also resampled.
(\textbf{C}) The state in one of the belief particles, $b^t_{\text{red},k}$ (top) and $b^{t+1}_{\text{red},k}$ (bottom). The particle is updated with ground truth properties blue object at step $t+1$. The properties of the pink object are resampled at step $t+1$ since its believed position in step $t$ conflicts with the observation at step $t+1$.}
\label{fig:belief_update}
\end{figure}

Figure~\ref{fig:belief_update} depicts an example of how an agent updates its belief from step $t$ to step $t+1$ based on its observation at step $t+1$.

\section{Model Implementation Details}
\setcounter{figure}{0}
\setcounter{table}{0}
\setcounter{equation}{0}

\subsection{SocialGNN}

The original SocialGNN model \citep{malik2023relational}  was designed to classify the social relationship between two agents (friendly, neutral, or adversarial) from visuospatial input. The model represents each video frame as a graph with nodes for agents and objects connected by bidirectional edges when entities are in physical contact. Landmark and wall coordinates were previously appended as global context features to each edge rather than represented as graph nodes. A single linear classification head operating on the final LSTM state produced a 3-class relation prediction. In the present work, we extended this architecture to jointly infer each agent's goal alongside the social relation, requiring modifications to both the graph representation and the output structure.

To represent goal-relevant dynamics, we extended the input graph by adding the landmarks in the scene as explicit graph nodes. In the original model, landmarks provided no direct relational signal between agents and landmarks. Because the goals in the PHASE dataset involve directed navigation toward or away from a landmark, or toward the other agent, making landmarks graph nodes allows the model to learn relational representations that are directly diagnostic of goal pursuit. Each landmark node follows the same feature vector structure as entity nodes, with velocity and angle set to zero as they are stationary. Correspondingly, we added a new class of bidirectional edges connecting each entity to any landmark it is in proximity to, analogous to the existing contact-based entity--entity edges. This structural expansion allows the model to capture the relative spatial dynamics necessary for goal recognition.

The PHASE dataset encodes each agent's goal as one of several discrete types, which can be categorized into three functional groups: (1) moving an object to a specific landmark, (2) navigating the agent itself toward or away from a landmark, and (3) pursuing a social interaction goal directed toward the other agent with a specific valence. The ground-truth label for each video is represented as a concatenated vector encoding the goal for each agent. At inference time, softmax is applied separately over each agent's goal partition to produce independent probability distributions over the possible intents.

We extended the $SocialGNN_{E}$ architecture with a second output head to enable joint prediction of goals and relations. The shared backbone remains identical to the original. The final LSTM state is passed in parallel to two linear classifiers: a relation head (3-class softmax) and a goal head that predicts the intents for both agents. The model is trained end-to-end by minimizing a combined loss $\mathcal{L} = \mathcal{L}_{\text{relation}} + \mathcal{L}_{\text{goal}} + \lambda \|\theta\|^2$, where both $\mathcal{L}_{\text{relation}}$ and $\mathcal{L}_{\text{goal}}$ are cross-entropy losses. Class weights are applied to the relation loss to upweight the underrepresented neutral class. This joint training objective encourages the shared representation to simultaneously capture features diagnostic of both social relations and individual agent goals.

\subsection{SIMPLE}

\subsubsection{Model parameters}

For SIMPLE model, we use the following parameters:

\begin{itemize}
    \item number of iterations: $L=6$
    \item number of particles/hypotheses: $M=15$
    \item scaling factor: $\eta=0.1$
    \item likelihood parameter $\beta=0.05$
    \item time interval for location estimation: $\Delta T=10$
\end{itemize}

\subsubsection{Bottom-up proposal for hypotheses}
We devise a bottom-up proposal based on heuristics extracted from observed trajectories within a time interval $S^{t_1:t_2}$, i.e., $h \sim Q(h|S^{t_1:t_2})$. In this work, the proposal distribution is decomposed into separate terms for proposing goals ($g_i$, $g_j$), social utility weights ($\alpha_{ij}$, $\alpha_{ji}$), and strengths ($f_i$, $f_j$), respectively, i.e.,
\begin{equation}
\begin{array}{ll}
    Q(h|S^{t_1:t_2}) =& Q(g_i|S^{t_1:t_2})Q(g_j|S^{t_1:t_2})\\
    &\cdot Q(\alpha_{ij}, \alpha_{ji}|S^{t_1:t_2})\\
    &\cdot Q(f_i|S^{t_1:t_2})Q(f_j|S^{t_1:t_2}).
\end{array}
\end{equation}

We define the goal proposal distribution for an agent by
\begin{equation}
\begin{array}{ll}
Q(g | S^{t_1:t_2}) &\propto e^{\gamma ||s_i^{t_2} - s_g||_2}e^{\gamma (||s_i^{t_2} - s_g||_2 - |s_i^{t_1} - s_g||_2)}\\
&\propto e^{\gamma (2||s_i^{t_2} - s_g||_2 - |s_i^{t_1} - s_g||_2)},
\end{array}
\end{equation}
where $\gamma = 10$ is a constant weight. Intuitively, if the trajectories have demonstrated either achievement at the end of the period ($t_2$) or progress towards a goal during the period (from $t_1$ to $t_2$), then that goal is likely to be the true goal. For the social utility weights, we first randomly select $u \in \{-1, 0, 1\}$. If $u = 0$, we set both $\alpha_{ij}$ and $\alpha_{ji}$ to be zero; if $u\in\{-1,1\}$, we randomly select either $\alpha_{ij}$ or $\alpha_{ji}$, and set it to be $u$ while setting the other one to be zero. This is essentially assuming that there will be at most one agent pursuing a social goal in a social event. For the strengths, we train a 2-layer MLP (64-dim for each layer) using training data in PHASE to estimate the maximum forces that agents can exert.

\subsection{Gemini 2.5 Pro}
For Gemini 2.5 Pro, we use the default model configuration (default model temperature and thinking budget). The following prompt is used for the model.

\begin{tcolorbox}[title=Agent Interaction Prompt, colback=gray!5, colframe=blue!75!black]
You are watching an animation of two agents acting in a 2D environment. Each animated video has the following:\\
* Entities: creatures (trapezoids) or objects (circles). Entities can be of different colors and sizes. Bigger entities are heavier. \\
* Landmarks: colored squares that creatures may try to get to or move objects to.\\
* Walls: black lines, creatures and objects can't move through walls.
\\
The agent's objective is to move to a landmark, move an object to a landmark, move close to the other agent, or hinder a particular goal (e.g., keep an agent or an object away from a landmark). By permuting the objects and landmarks, we have 36 possible goals in total for the goal space.

The agent's relation can be friendly (collaborating on a goal), neutral  (pursuing independent goals) or adversarial (e.g. hindering or competing with other agent).

Your task is to classify each agent's goal and relations. You will be given the full goal space and possible relation labels and you will output a probability distribution of top 10 goals for each agent, and a probability distribution of possible agent relations.

1. red\_agent\_goals: exactly 10 goals for the red agent. Dictionary mapping goal string -> probability. Choose only from this goal space:
\{goals\_str\}

2. green\_agent\_goals: exactly 10 goals for the green agent. Same format, same goal space.

3. relation: distribution over the relationship. Dictionary with exactly three keys "Adversarial", "Friendly", "Neutral"; values sum to 1.0.

Your response must be ONLY valid JSON---one object with exactly these 3 keys, no other text:

\begin{verbatim}
{
  "red_agent_goals": {
    "<goal from list>": 0.12,
    "<goal from list>": 0.11,
    ...exactly 10 entries, values sum to 1.0
  },
  "green_agent_goals": {
    "<goal from list>": 0.15,
    ...exactly 10 entries, values sum to 1.0
  },
  "relation": {
    "Adversarial": 0.1,
    "Friendly": 0.7,
    "Neutral": 0.2
  }
}
\end{verbatim}
\end{tcolorbox}

\subsection{Predicate Inverse Planning}
The \textbf{predicate inverse planning} model infers agents' goals and their social relations from observed trajectories by (1) abstracting continuous states into symbolic predicates, and (2) inverting a forward planning process over those symbolic states. It does \emph{not} use physical dynamics; it operates only on a discrete, predicate-based representation of the scene over time.

Let the continuous state at time $t$ be $\mathbf{s}_t \in \mathcal{S}$ (e.g., positions and velocities of agents and objects). The model first maps $\mathbf{s}_t$ to a set of Boolean predicates evaluated at thresholds, yielding a \textbf{symbolic state} $\phi_t$. So at each $t$, we have a vector (or set) of predicate truth values:
\[
  \phi_t = \bigl\{ \mathrm{ON}(e,\ell), \mathrm{TOUCH}(e_1,e_2), \mathrm{ATTACH}(a,o), \mathrm{CLOSE}(\cdot,\cdot), \ldots \bigr\}.
\]
The observed trajectory is then represented as a \textbf{symbolic trajectory} $\tau_\phi = (\phi_0, \phi_1, \ldots, \phi_T)$.

Assume a (symbolic) forward model: given the goal $g_i$, the agent is assumed to produce a sequence of actions that, under a simplified symbolic dynamics, leads to predicate trajectories that satisfy or progress toward the goal. Let $\mathbb{P}(\tau_\phi \mid g_i)$ denote the probability of the observed symbolic trajectory $\tau_\phi$ given that agent $i$ has goal $g_i$ (and optionally given the other agent's goal or a relation prior). This is typically implemented by forward-simulating or scoring candidate goals against $\tau_\phi$ (e.g., by checking which goal best ``explains'' the observed ON/TOUCH/ATTACH/CLOSE pattern over time).

Inverse planning inverts this forward model under a prior over goals $P(g_i)$ (often uniform over $\mathcal{G}$):
\begin{equation}
  P(g_i \mid \tau_\phi) \propto P(\tau_\phi \mid g_i)\, P(g_i).
  \label{eq:inverse_appendix}
\end{equation}

Unlike physics-based models (e.g., SIMPLE), this model does \emph{not} use forces, masses, or continuous dynamics. It only sees which predicates hold at each time. As a result it can do well when relations are discernible from coarse symbolic structure (e.g., Independent when each agent moves toward different objects/landmarks), but it struggles when fine-grained physics is required (e.g., whether an agent can actually push an object to a landmark, or whether two agents are opposing vs.\ cooperating on the same object), leading to lower goal accuracy than physically grounded inverse planning.

\section{Additional Experiment Details}
\setcounter{figure}{0}
\setcounter{table}{0}
\setcounter{equation}{0}

\subsection{Goals in PHASE}

There are 36 possible goals in PHASE, which are listed below:

\begin{verbatim}
"Get blue item to blue lm",
"Get blue item to green lm",
"Get blue item to red lm",
"Get blue item to yellow lm",
"Get blue item away from blue lm",
"Get blue item away from green lm",
"Get blue item away from red lm",
"Get blue item away from yellow lm",
"Get green agent to blue lm",
"Get green agent to green lm",
"Get green agent to red lm",
"Get green agent to yellow lm",
"Get green agent away from blue lm",
"Get green agent away from green lm",
"Get green agent away from red lm",
"Get green agent away from yellow lm",
"Get pink item to blue lm",
"Get pink item to green lm",
"Get pink item to red lm",
"Get pink item to yellow lm",
"Get pink item away from blue lm",
"Get pink item away from green lm",
"Get pink item away from red lm",
"Get pink item away from yellow lm",
"Get red agent to blue lm",
"Get red agent to green lm",
"Get red agent to red lm",
"Get red agent to yellow lm",
"Get red agent away from blue lm",
"Get red agent away from green lm",
"Get red agent away from red lm",
"Get red agent away from yellow lm",
"get away from green agent",
"get away from red agent",
"get to green agent",
"get to red agent"
\end{verbatim}

\subsection{Accuracy}
For human accuracy, we compute the per participant accuracy on goal/relation classification and average across all participants. Suppose N participants attempt the experiment. The accuracy is computed by

\begin{equation}
\text{Goal Accuracy} = \frac{1}{N} \sum_{i} \left( \frac{\text{Number of correct goal classifications by participant i} }{\text{Total number of goal classifications by participant i}} \right)
\end{equation}

The accuracy of models are computed by taking the option with the highest assigned probability and comparing against the ground-truth label.

\subsection{Correlation}
To compute model/human or human split-half correlations, we first normalize the human ratings. Each video has 2 goals and 1 relation. For each goal, we get a distribution over 36 goal labels. For video $v$, the probability of agent $i$ having goal $g$ is defined as

\begin{equation}
P(\text{agent $i$ has goal $g$ in video $v$}) = \frac{\text{total number of people chose $g$ for agent $i$ in video $v$}}{\text{total number of people who rated the goal of agent $i$ in video $v$}}
\end{equation}

Since we have 100 videos, 2 goal tasks per video with 36 possible goals per task, 1 relation task per video with 3 possible relation labels per video, this produces 7200 probability measures for goals and 300 probability measures for relations. We flatten these into one goal vector of dimension 7200 and one relation vector of dimension 300, and then compute the correlation between human subgroups or humans and models, where the same dimension vector can be extracted from model distribution over goals and relation labels.

\end{document}